# Excimer-Suppressed and Oxygen-Tolerant Photophysics of 'Arm-like' Substituted Pyrene Derivatives

Wenlong Li,[†] Stephen Awuku,[†] Jenna N. Merk,[‡] Marc R. MacKinnon,[‡] and Amy L. Stevens[*,†,b]

*Corresponding author

[†]Department of Chemistry, University of Saskatchewan, Saskatoon, SK, Canada S7N 5C9

[b]Centre for Quantum Topology and Its Applications (quanTA), University of Saskatchewan, SK, Canada S7N 5E6

[‡]Department of Chemistry and Biochemistry, University of Regina, Regina, SK, Canada S4S 0A2

## ABSTRACT:

Pyrene-functionalized materials are extensively employed in photoluminescent applications, owing to their extended π-conjugation and favorable photophysical properties. However, their luminescent performance is often attenuated by π-π stacking-driven excimer formation and molecular oxygen quenching. To mitigate these undesirable effects, a novel class of 7-*tert*-butylpyren-2-ol derivatives with extended 'arm-like' substituents at the 1,3-positions have been synthesized and their luminescent properties in solution have been thoroughly investigated. While the 2- and 7-positions of the pyrene core are frequently modified with hydroxyl and *tert*-butyl groups, this work presents the first introduction of 'arm-like' substituents at the 1,3-positions. The stretched-out 'arm-like' substituents not only introduce steric bulk to suppress excimer formation but also change the symmetry class of pyrene and modulate electron density at its 1,2,3,7-positions. These effects tune pyrene's energy levels and result in moderate (0.4) to high (0.7) fluorescence quantum yields and shorter-lived fluorescence lifetimes ranging from ca. 20 to 40 ns. These shorter lifetimes lead to a reduction of the pyrene derivatives' susceptibility to energy scavenging by molecular oxygen. In addition, the specific form of the 'arms' are important. Alkyl-containing arms and alkenyl-containing arms exhibit different decay pathways, which is reflected by their disparate nonradiative rates. Thus, the introduction of 'arm-like' modifications represents a promising approach to modulate the photophysical behaviour of an important class of polycyclic aromatic hydrocarbons, highlighting their applicability in next-generation electronic and optoelectronic systems.

# TOC Graphic

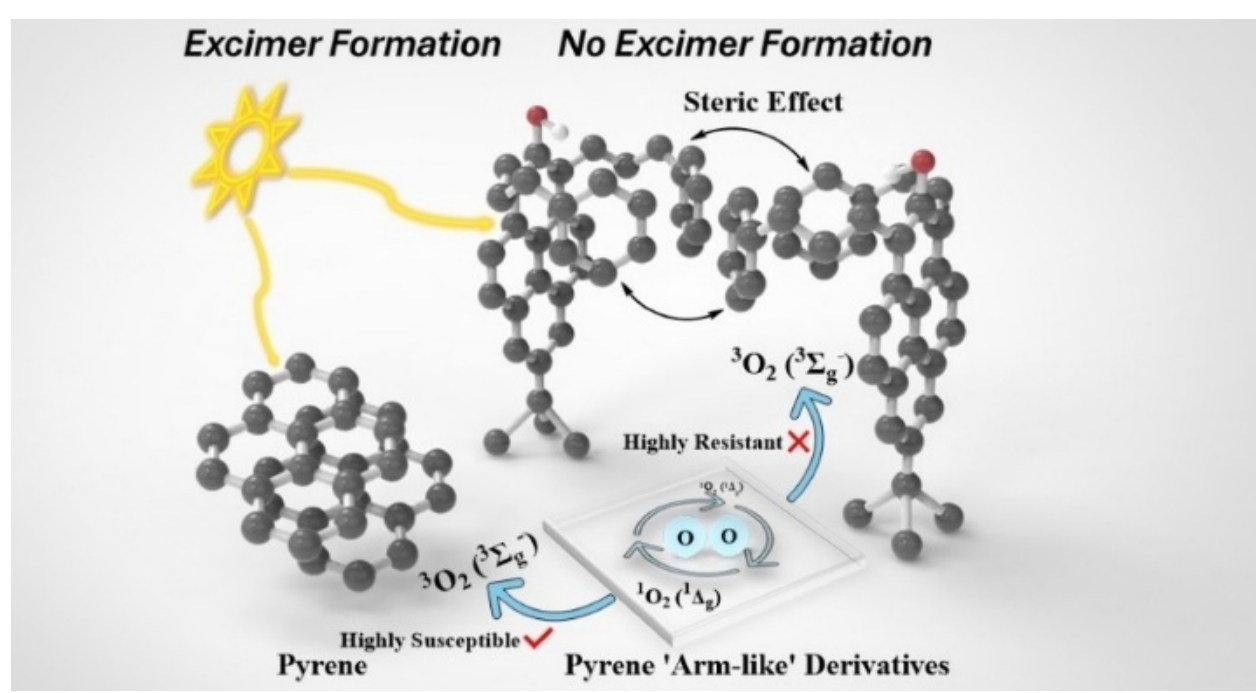

## INTRODUCTION

Pyrene is an important member of the family of polycyclic aromatic hydrocarbons. Pyrene and its derivatives have been extensively researched due to their unique and excellent photophysical and electronic properties in solution and in the solid state. Pyrene's intriguing photophysical properties are due to its high symmetry ($D_{2h}$ point group), which includes an inversion centre. The extraordinarily high photostability of pyrene and the ease with which its molecular structure can be modified have led to the widespread adoption of pyrene-based materials in diverse applications. These applications include organic light-emitting diodes (OLEDs),[1–3] organic field-effect transistors (OFETs),[4,5] organic photovoltaic cells (OPVs),[6] and dye-sensitized solar cells (DSSCs).[7,8] Moreover, recent advances have highlighted the potential of pyrene-based materials as light-harvesting antenna systems.[9] In addition, pyrene and pyrene derivatives possess favourably high fluorescence quantum yields (QYs) and long fluorescence lifetimes, which enable their use as bioimagers,[10,11] biosensors,[12] and fluorescent probes.[13–15] Another unique feature of pyrenes is that their excited-state π-π stacking interactions lead to the formation of excited-state dimers or 'excimers'. This excimer-forming ability has been exploited to produce extremely sensitive probes for the folding of biomolecules in live cells.[16,17] More recently, the concentration- and distance-dependent nature of excimers has led to the use of pyrene-based molecules as drug delivery[18] and environmental monitoring[19] tools.

Although excimer formation is an indispensable photophysical signature that is exploited in many technologies, it is an undesirable byproduct in devices that have light-harvesting or energy-conversion[2,3,20,21] properties. The formation of excimers in pyrene and pyrene derivatives leads to significantly quenched monomeric fluorescence emission, with a concomitant shortening of their fluorescence lifetimes. In addition, excimers have a propensity to act as trap states that impede key photophysical processes such as energy transfer and exciton diffusion, which are crucial mechanisms for efficient optoelectronic and photonic applications.[20] Dissolved molecular oxygen has a similar effect to that of excimer formation, although its fluorescence-quenching mechanism stems from intersystem crossing (ISC).[22] For example, benzo[a]pyrene in acetonitrile undergoes ISC from its singlet states to its triplet states upon light irradiation, yielding singlet oxygen or superoxide radicals via energy or electron transfer processes.[22] Therefore, removing dissolved oxygen from a solution significantly alters the fluorescence emission behaviour of pyrene derivatives. After degassing, the fluorescence lifetimes of these derivatives in most solvents increase from 1 – 10 ns to 50 – 80 ns and modest to high fluorescence quantum yields are achieved.[3,21]

Various strategies have been employed to tune the structure of pyrenes to prevent excimer formation. For example, Crawford et al. have substituted functional groups, such as R = -$O(CH)_{12}Br$, -C≡C(TMS), -$C_6H_4$-4-$CO_2Me$, at the 1- or 2,7-positions on the pyrene core; see Scheme 1. These elongated-chain substituents prevent excimer formation due to an unspecified form of steric hindrance, which is further aided by low sample concentrations ($10^{-6}$ to $10^{-5}$ M).[21] Xie et al. have employed bulky groups such as *tert*-butyl at the 7-position on the pyrene core to reduce excimer formation and the strength of π-π stacking in both solution and the solid state. Notably, bulky groups incorporated at the 1,3-positions, such as triphenylamine and tetraphenylethylene, suppress intermolecular π-π stacking interactions.[2] Haedler et al. modified the substituent at the 3-position on the pyrene core. Their long chain substituent at this position contained either a methylene-ester or methylene-amide linker, which was followed by a benzene or 4-*tert*-butyl benzene group. These modifications were found to reduce excimer formation, which becomes significant at high concentrations.[23]

To avoid excimer formation and to decrease the effect of oxygen quenching on the monomeric fluorescence emission, five different 1,3-extended-chain substituted 7-*tert*-butylpyren-2-ol pyrene derivatives (Scheme 1) have been synthesized and spectroscopically investigated in this work. The bulky *tert*-butyl group is substituted at the 7-position on the pyrene core to sterically hinder excimer formation and π-π stacking interactions. Synthetically, the *tert*-butyl also serves as a steric barrier, promoting further functionalization at the other end of the molecule. At the 2-position, a hydroxyl group is incorporated to tune the photophysics for OLED applications. However, this is believed to decrease the fluorescence quantum yield of pyrenes because of the resulting strong intermolecular hydrogen bonding interactions from the hydroxyl group. The methylation of the hydroxyl group at the 2-position can improve the fluorescence quantum yield.[2] Recently, the hydroxyl group incorporated in the pyrene core has been shown to further suppress π-π stacking interactions and restrict molecular motions due to hydrogen bonding. Therefore, adding this substituent may enhance pyrene's luminescent properties.[3] However, the most distinctive feature of this class of pyrene derivatives is the incorporation of elongated, 'arm-like' substituents at the 1,3-positions of the pyrene. These substituents are bulky, being approximately 0.5 to 1.0 nm in length, which can transiently or partially obstruct the pyrene core. If this obstruction occurs, we believe there will be an additional spatial hindrance to excimer formation. Furthermore, the incorporation of 'arm-like' substituents is anticipated to significantly alter the photophysical properties of pyrene compared to those observed in their absence, which will also alter the energy-scavenging ability of oxygen.

In this paper, we characterize the photophysical properties of these novel 1,3-extended-chain substituted 7-*tert*-butylpyren-2-ol derivatives by steady-state absorption, steady-state fluorescence excitation and emission, and fluorescence lifetime measurements. From this data we determine their fluorescence quantum yields and their excitation decay parameters. By comparing the photophysics of the 'parent' pyrene and this class of pyrene derivatives we highlight their potential significance for electronic and optoelectronic devices. We also report that the innovative 1,3-position 'arm-like' substituents hinder excimer formation and the altered fluorescence lifetimes mitigate issues related to oxygen scavenging. This introduces a novel synthetic strategy to address the challenges posed by excimer formation and oxygen-related fluorescence quenching in pyrenes.

**Scheme 1. Structures for Pyrene and the Synthesized and Investigated Pyrene Derivatives. The International Union of Pure and Applied Chemistry (IUPAC) name and our abbreviation are given for each compound.**

Pyrene
Py

7-*tert*-butyl-1,3-(diprenyl)pyren-2-ol
$(Pren)_2$-Py

7-*tert*-butyl-1,3-(digeranyl)pyren-2-ol
$(Gera(en)_2)_2$-Py

7-*tert*-butyl-1,3-di(3-methylpropyl)-pyren-2-ol
$(MeProp)_2$-Py

7-*tert*-butyl-1,3-di(3-phenylpropyl)-pyren-2-ol
$(PhProp)_2$-Py

7-*tert*-butyl-1,3-di(3-phenylpropen-2-yl)-pyren-2-ol
$(PhPren)_2$-Py

## EXPERIMENTAL SECTION

**Materials.** Pyrene (98% purity) was purchased from Sigma-Aldrich and used without further purification. This purity was checked by comparing the experimentally determined melting temperature (using a DigiMelt MPA160) with the reported one in the CAS SciFinder database and the one provided by the vendor (Sigma-Aldrich). These melting points of pyrene were consistent within experimental error. The synthetic methods and characterization details of the investigated pyrene derivatives, shown in Scheme 1, are provided in the Supporting Information (SI). Solvents were purchased from Sigma-Aldrich and Fisher Chemicals and used as received. The solvents used were spectroscopic grade toluene, tetrahydrofuran (THF), dimethyl sulfoxide (DMSO), cyclohexane, and methanol. Toluene was used preferentially throughout the study due to its superior solubility of pyrene and the pyrene derivatives. However, other listed solvents were used as comparisons when the examined compounds were fully soluble.

**Techniques.** All experiments were performed at room temperature. The repeated process of freeze-pump-thaw on a vacuum line (5 cycles per sample) was used to remove any dissolved oxygen in the solvents for the fluorescence quantum yield and fluorescence lifetime measurements. All degassed measurements were carried out within 1 hour after the completion of each degassing cycle and the absorption spectrum of each sample was measured before and after the degassing procedure to ensure that the sample remained undamaged.

To prepare the solutions, dry pyrene and its derivatives were weighed by a precise analytical balance and dissolved in spectroscopic grade toluene (or other solvents) to make $10^{-3}$ M (or 250 μM) stock solutions. Afterwards, the stock solutions were stirred for approximately 30 minutes at room temperature and then sonicated to achieve complete dissolution and homogeneity. A series dilution technique was performed to obtain accurate concentrations ranging from $10^{-3}$ M (or 250 μM) to $10^{-6}$ M as required. To avoid the photodegradation of these samples and evaporation of their solvents, the prepared solutions were placed in glass vials (used for light-sensitive materials) and then stored in a refrigerator. Before being measured, the solutions were taken from the refrigerator to a dark and cool place until they returned to room temperature. The following spectroscopic measurements were repeated 5 times for each sample.

The UV-visible absorption spectra and molar absorption coefficients were obtained using either a Cary-6000i dual-beam spectrophotometer, an Edinburgh Instruments spectrofluorometer FS5, or a Cary-3500 multicell UV-vis spectrophotometer. To circumvent the saturation effects of the instruments, various quartz absorbance cuvettes with pathlengths from 10 mm to 0.1 mm were used for these measurements.

The photostability of pyrene and its derivatives were examined by irradiating 3 mL of their vigorously stirred solutions by a 355 nm continuum-wave laser (average power of ca. 8 mW) for a certain length of time. Their absorption spectra were examined to determine the loss of absorbers over time. This was quantified by calculating photodegradation quantum yields. The pyrene derivatives were found to be photo-unstable under these conditions. Thus, the samples were further examined by irradiating with a 1000-fold less powerful 340 ± 10 nm (337 nm) diode laser (average power of ca. 1 μW) contained in the Edinburgh Instruments spectrofluorometer

FS5. No changes in peak intensity or wavelengths were observed in the UV-visible absorption spectra for all investigated samples over a 2-hour period.

The fluorescence excitation and steady-state fluorescence emission spectra were measured using the Edinburgh Instruments spectrofluorometer FS5. The spectrofluorometer FS5 is equipped with a 150 W xenon arc lamp as one of the excitation sources, and the detector is R928P (Hamamatsu). The fluorescence was collected at 90 degrees with respect to the excitation beam to minimize scatter from the excitation source. A single set of triangular cuvettes or 10 mm × 2 mm fluorescence cuvettes were employed to measure the fluorescence emission for samples in an aerated environment at all concentrations. Degassed dilute samples ($10^{-6}$ to $10^{-5}$ M) were used for fluorescence measurements to eliminate quenching resulting from aggregates and dissolved oxygen. The same set of 10 mm × 2 mm fluorescence degassing cuvettes were employed to compare different measurements (absorption and emission) simultaneously in aerated and degassed solutions throughout the experiments.

Time-resolved fluorescence spectra were collected using the time-correlated single photon counting (TCSPC) technique in the Edinburgh Instruments spectrofluorometer FS5. The dilute degassed samples were excited by the FS5's 337 nm picosecond pulsed laser diode (EPLED-UV, 340 ± 10 nm, 2 MHz pulse train, and 14 ± 4 nm (FWHM)). Their fluorescence was detected at wavelengths of ca. 370 nm for pyrene and ca. 387 to 395 nm for the pyrene derivatives, which corresponded to the near maximum fluorescence intensity in the fluorescence emission band for each sample. Peak-channel counts varied with concentration, ranging from approximately 1700 to 5100 for pyrene and from 7300 to 18500 for the pyrene derivatives. The instrument response function (IRF) was obtained using a dilute Ludox scatterer at the excitation wavelength, yielding a typical FWHM of ca. 1.2 ns. Following subtraction of the background, the fluorescence lifetimes were acquired by iterative reconvolution of the IRF as one-component or two-component exponential decay functions. For each measurement, the goodness of fit was evaluated by a reduced $\chi^2$ value and the weighted residual plot. The best fit was not obtained until the value of $\chi^2$ was minimized and there were an even number of data points either side of the regression line.

Fluorescence quantum yields were determined using degassed dilute samples ($10^{-6}$ to $10^{-5}$ M), with absorbances less than 0.1 in a 10 mm × 10 mm absorption cuvette at the excitation wavelength. This low concentration circumvents artifacts from self-quenching and reabsorption effects. The excitation wavelength was set at 330 nm, and emission was collected from 340 to 650 nm. Subsequently, the fluorescence emission spectra were corrected for background solvent, followed by the correction of different absorbances at the excitation wavelength. Pyrene in toluene was used as a fluorescence quantum yield standard and the fluorescence quantum yields of the pyrene derivatives were calculated using the following equation.

$$\Phi_{\text{sample}} = \Phi_{\text{standard}} \times \frac{A_{\text{sample}}}{A_{\text{standard}}} \times (\frac{n_{\text{sample}}}{n_{\text{standard}}})^2$$

where $\Phi_{\text{sample}}$ and $\Phi_{\text{standard}}$ are the fluorescence quantum yields of the sample of interest and that of the standard, respectively. $A_{\text{sample}}$ and $A_{\text{standard}}$ are the integrated areas of the fully corrected fluorescence emission spectra of the sample of interest and that of the standard, respectively. $n_{\text{sample}}$ and $n_{\text{standard}}$ represent the refractive indices of the solvent used for the sample and the standard at the excitation wavelength, respectively. The fluorescence quantum yield of pyrene in degassed toluene was calculated from the well-known fluorescence quantum yield of pyrene in degassed cyclohexane using the same equation.

## RESULTS AND DISCUSSION

**Pyrene.** The photophysics of pyrene in toluene will first be discussed prior to examining the substituent effects. The photophysics of pyrene in some common solvents such as THF, chloroform, ethanol, and cyclohexane, most importantly, is well-studied.[24–26] Toluene solutions of pyrene have been less extensively studied. However, toluene is the sole solvent capable of completely dissolving all pyrene derivatives examined in this study. Furthermore, toluene is an aromatic hydrocarbon capable of π–π stacking with pyrene and its derivatives. This could lead to spurious spectral artifacts, and so a concentration study of pyrene in toluene was performed.

**Absorption Spectroscopy.** The absorption spectrum of pyrene in toluene consists of two bands (Figure 1A), with a toluene UV cut-off at ca. 285 nm. There is a weak absorption band from 350 nm to 390 nm and a strong band from 285 nm to 350 nm. Both bands have strong vibronic character. The origin (0-0) transition peak of the $S_0 \rightarrow S_1$ transition is at 372.9 nm. This transition has a molar absorption coefficient (ε) of 300 $M^{-1}$ $cm^{-1}$ for concentrations larger than 1 μM, and an ε value of 1200 $M^{-1}$ $cm^{-1}$ for concentrations equal to or less than 1 μM. As previously mentioned, pyrene exhibits $D_{2h}$ symmetry. The centrosymmetric nature of pyrene results in a symmetry-forbidden $S_0 \rightarrow S_1$ transition, as dictated by the orbital selection rule.[20] This is the reason for the low molar absorption observed in the $S_0 \rightarrow S_1$ transition band. The origin (0-0) transition peak of the strongly-allowed band, the $S_0 \rightarrow S_2$ transition, is at 337.8 nm with an ε of 52000 $M^{-1}$ $cm^{-1}$ for concentrations larger than 1 μM. Again, at concentrations equal to or less than 1 μM a higher ε value of 71000 $M^{-1}$ $cm^{-1}$ is seen. Previous studies are in line with our observations of the $S_0 \rightarrow S_1$ and $S_0 \rightarrow S_2$ transition maxima[21] and their relative intensities.[27,28] The strengths of the absorption bands have been shown to reflect the intrinsic allowedness of electronic transitions for unsubstituted pyrene, regardless of the solvent environment.[27,28] The distinctive ε values observed above and below a concentration of 1 μM for pyrene in toluene are likely due to a trace amount of ground-state dimerized pyrene or pre-associated pyrene formed via π-π stacking interactions. This is being investigated currently.

**Fluorescence Spectroscopy.** The fluorescence spectrum of dilute pyrene ($10^{-6}$ to $10^{-5}$ M) in toluene, in Figure 1B, shows no excimer formation. However, the spectral signature of excimers begins to emerge within the concentration range of $10^{-5}$ to $10^{-4}$ M, becoming notably prominent starting at a concentration of ca. 30 μM. The pyrene monomer exhibits a vibronically-structured fluorescence band from 350 to 425 nm, with a maximum of 374 nm at all concentrations. The fluorescence emission of pyrene excimers is redshifted compared to the monomer emission, with a maximum at ca. 475 nm, and is broad and structureless. The fluorescence intensity of this band increases as pyrene becomes more concentrated, which is the telltale signature of excimers in solution.[29,30] The findings from steady-state absorption and fluorescence measurements of pyrene in toluene indicate that there are minimal sample-solvent interactions between pyrene and toluene. For the pyrene derivatives, the presence of the bulky substituents further suppresses any potential solvent effects, effectively shielding the chromophores from interactions with toluene.

**A**

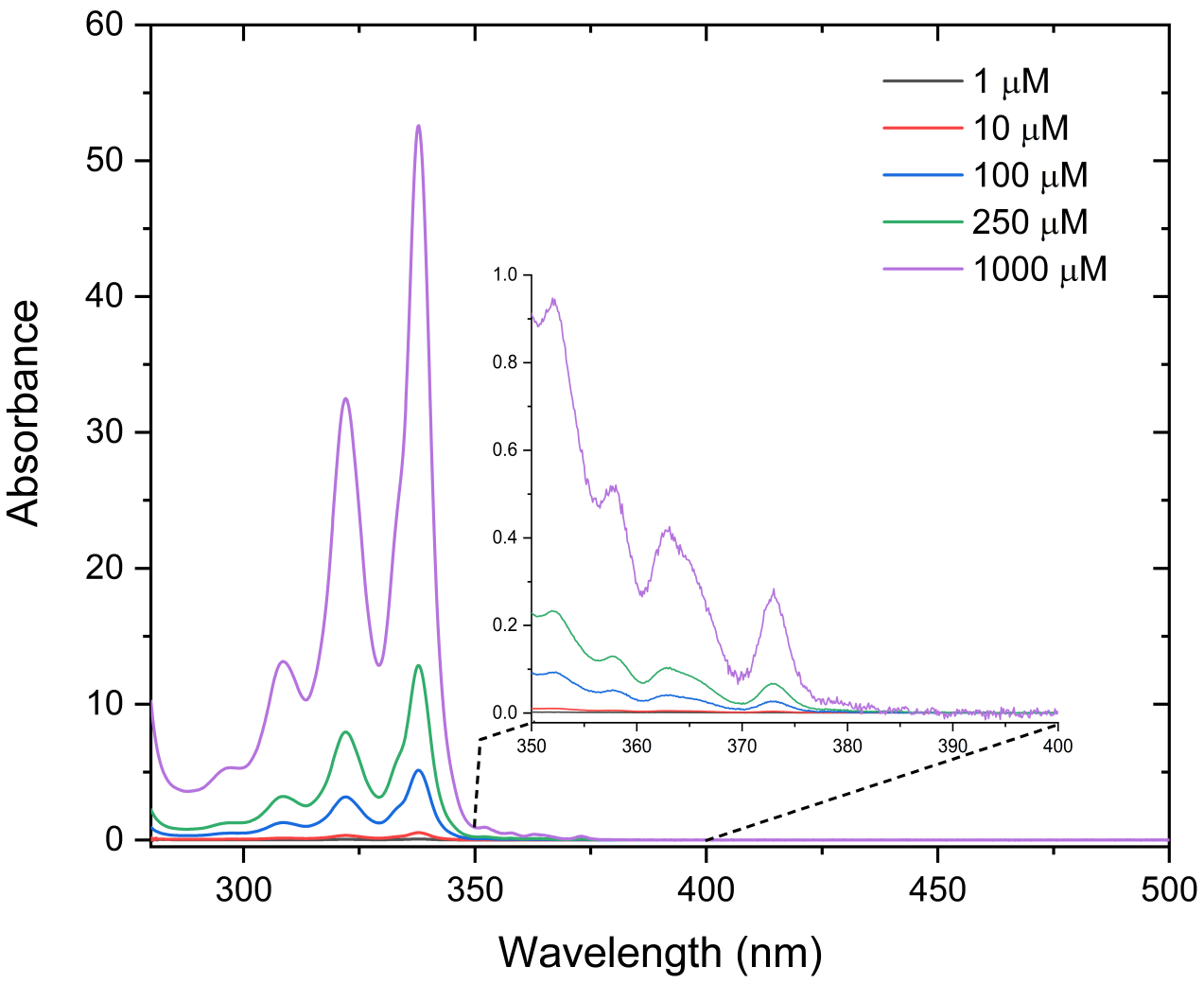

**B**

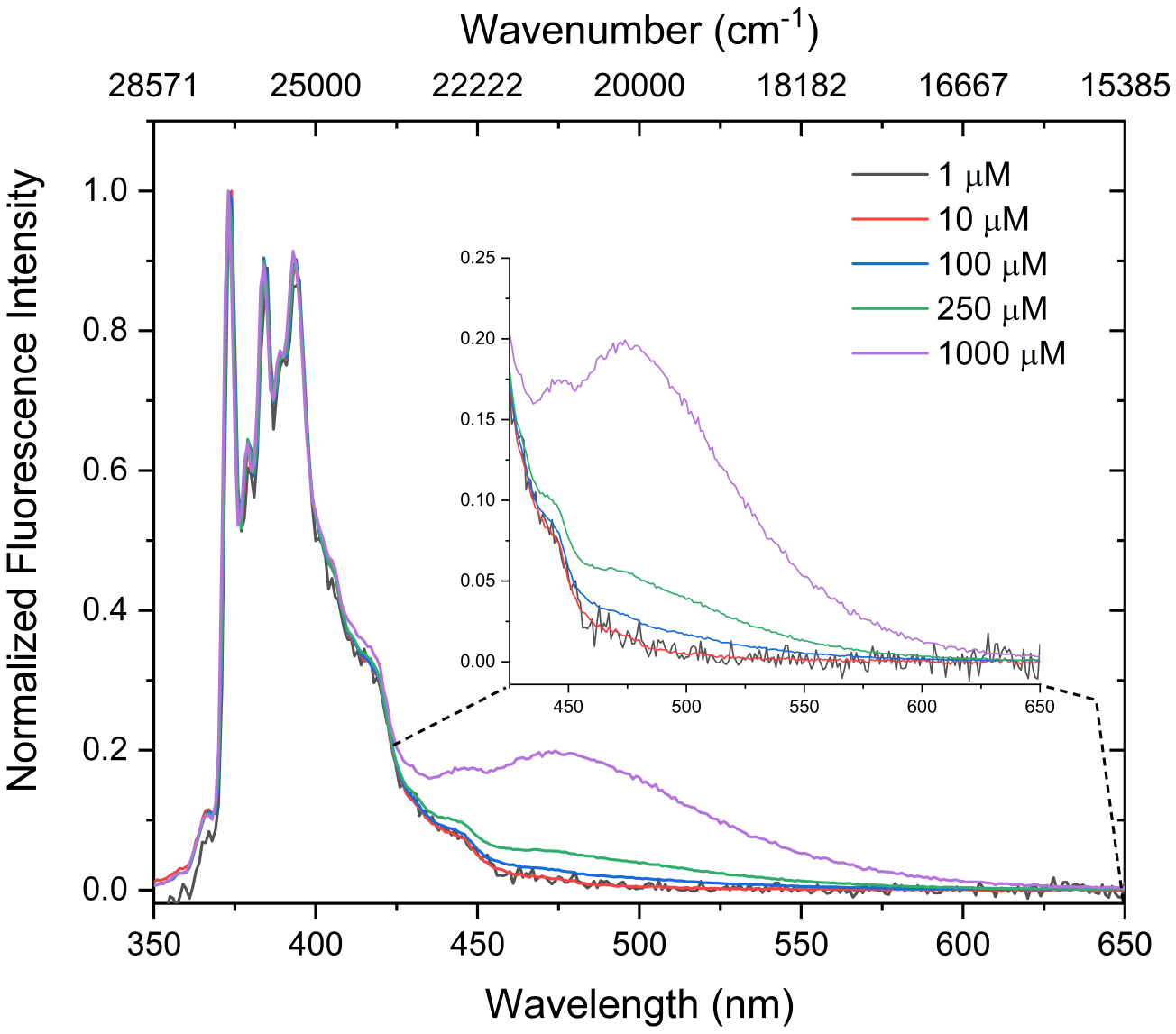

**Figure 1.** (A) Concentration dependent studies of pyrene in toluene by absorption spectroscopy in an aerated environment at room temperature. The inset shows the expanded view of the forbidden transition between the electronic ground state ($S_0$) and the electronic first excited state ($S_1$). (B) Normalized fluorescence emission spectra of pyrene at various concentrations in toluene in an aerated environment at room temperature. Excitation is at 330 nm. Each spectrum

was normalized at its maximum intensity. The inset provides an expanded view of the excimer fluorescence.

**'Arm-like' Pyrene Derivatives.** To assess the photoluminescent properties of these 'arm-like' pyrene derivatives, their absorption, fluorescence emission and excitation, fluorescence quantum yields, and lifetimes are measured. From this data, their kinetic parameters are then calculated. To ensure no specific solvent-solute complexes were formed, one of the derivatives, $(Gera(en)_2)_2$-Py, was examined in a range of solvents with different polarities. Consistent with expectations, the absorption spectra show minimal shifts and no evidence of new absorption features. Therefore, there are no charge transfer states or π–π stacking interactions between the derivatives and toluene; see Figure S1A in the SI. To ensure the photostability of the 'arm-like' pyrene derivatives, they were subjected to high-power 355 nm CW laser irradiation, as presented in Figure S2 of the SI. For the 'arm-like' derivatives with alkenyl linkers at 1,3-positions such as $(PhPren)_2$-Py, the absorbance intensity of all transitions decreases significantly over time. In contrast, for derivatives with alkyl linkers at the 1,3-positions — such as $(PhProp)_2$-Py — a new and intense absorption band centered between 480 – 485 nm emerges, alongside an overall decline in absorbance intensity. These results indicate low photostability under CW high-power laser irradiation. However, the samples remain photostable and are well-suited for measurements using other commercial instruments (see Experimental section for details). The origin of this spectral evolution is not yet fully elucidated, but it is likely associated with photochemical processes, such as photo-oxidation of the pyrene core or pyrene radical formation.[31–33]

**Absorption Spectroscopy.** Figure 2 presents the key absorption spectral features of dilute pyrene and its derivatives observed in toluene. For the full-range absorption spectra between 280 to 800 nm recorded at high concentrations (250 μM), which clearly reveal weak transition bands at longer wavelengths due to the formation of ground-state aggregates, refer to Figure S3 in the SI. The key spectroscopic parameters are summarized in Table 1. For the $S_0 \rightarrow S_2$ transition, all 'arm-like' derivatives exhibit absorption features similar to those of unsubstituted pyrene, characterized by a distinct vibrational progression. The absorption maxima corresponding to the $S_0 \rightarrow S_2$ transition in all derivatives are redshifted by approximately 1350 – 1500 $cm^{-1}$ relative to unsubstituted pyrene. This redshift can be largely attributed to the synergistic effects of the hydroxyl group's electron-donating character and the increased electron density introduced by the 'arm-like' substituents. The derivatives with alkenyl linkers such as $(Pren)_2$-Py, $(Gera(en)_2)_2$-Py, and $(PhPren)_2$-Py exhibit the most redshifted $S_0 \rightarrow S_2$ transitions. This is due to the smaller HOMO-LUMO gap that exists due to the extra conjugation in their 'arms'. The substituents also affect the intensities of the absorption bands. All of the 'arm-like' derivatives exhibit ε values ranging from 22000 to 40000 $M^{-1}$ $cm^{-1}$ for the $S_0 \rightarrow S_2$ transition maxima, which are notably lower than that of pyrene, i.e. (52000 to 71000 $M^{-1}$ $cm^{-1}$).

A comprehensive explanation of the photophysical behavior of these pyrene derivatives requires simultaneous consideration of the $S_0 \rightarrow S_1$ and $S_0 \rightarrow S_2$ transitions in relation to the molecular orbital interactions involved. The $S_0 \rightarrow S_1$ transitions in the 'arm-like' pyrene derivatives are less forbidden, compared to that in pyrene, as evidenced by their increased ε values (Figure 2) and their excited state lifetimes (vide infra). This enhancement aligns with the literature in that substituents at the 2,7-positions significantly influence the $S_0 \rightarrow S_1$ transitions, which primarily involve HOMO - 1 and LUMO + 1 orbital interactions. In contrast, these substituents exert minimal impact on the $S_0 \rightarrow S_2$ transitions, which are mainly governed by HOMO and LUMO orbitals. This effect arises from the presence of both nodal and non-zero orbital contributions at the 2,7-positions in the molecular orbitals involved in the $S_0 \rightarrow S_2$ and $S_0 \rightarrow S_1$ transitions, respectively. Therefore, the hydroxyl group at the 2-position and the *tert*-butyl group at the 7-position on the pyrene core enhance the allowedness of the $S_0 \rightarrow S_1$ transitions. Additionally, substitutions at the 1- and 3- positions on the pyrene core have been found to significantly enhance the allowedness of both the $S_0 \rightarrow S_1$ and $S_0 \rightarrow S_2$ transitions. This conclusion is supported by a combination of experimental spectral features and theoretical density functional theory (DFT) calculations.[21] However, the less forbidden $S_0 \rightarrow S_1$ transitions are seen in this study, but not the $S_0 \rightarrow S_2$ transitions owing to a competing effect between the 1,3-subsituted and the *tert*-butyl groups. Collectively, the structural modifications of the derivatives result in a substantial increase in the transition probability.

Additionally, the role of symmetry breaking among these ‘arm-like’ derivatives cannot be underestimated. While unsubstituted pyrene possesses $D_{2h}$ symmetry, the substituent-induced modulation of the pyrene core reduces these ‘arm-like’ derivatives to $C_{2v}$ symmetry, which lacks an inversion center. According to the Laporte rule, electronic transitions that are forbidden in inversion center containing centrosymmetric molecules become allowed when molecular symmetry is broken,[34,35] which is the case for our ‘arm-like’ derivatives. In addition, symmetry breaking mixes the two singlet excited states, i.e. $S_1$ borrows oscillator strength from $S_2$.[36,37] This explains the intensification of the $S_1$ absorption transition and the decrease of the $S_2$ absorption transition. From this perspective, the forbidden $S_0 \rightarrow S_1$ transitions in unsubstituted pyrene become allowed in the ‘arm-like’ derivatives due to a combination of substituent effects and symmetry breaking.

Compared to unsubstituted pyrene, the $S_0 \rightarrow S_1$ absorption band in the ‘arm-like’ derivatives is also redshifted, as discussed for the $S_0 \rightarrow S_2$ transition. Moreover, the $S_0 \rightarrow S_1$ transitions in these ‘arm-like’ derivatives exhibit pronounced vibronic character. ‘Arm-like’ derivatives with alkenyl linkers such as $(Pren)_2$-Py, $(Gera(en)_2)_2$-Py, and $(PhPren)_2$-Py display comparable absorption profiles, with slight shifts in peak wavelengths. Owing to their similar molecular structures, $(Pren)_2$-Py and $(Gera(en)_2)_2$-Py show nearly identical absorption behaviour in the $S_0 \rightarrow S_1$ transitions. However, the intense peak centred at ca. 376 nm likely arises from contributions of both the $S_0 \rightarrow S_1$ and $S_0 \rightarrow S_2$ transitions. This phenomenon has been reported previously.[21] For highly conjugated 1-substituted pyrene systems, the spectral overlap between the $S_0 \rightarrow S_1$ and $S_0 \rightarrow S_2$ transitions causes them to be experimentally indistinguishable.[21] Due to the presence of alkenyl linkers in the ‘arm-like’ substituents, these derivatives also have extra low-frequency torsional modes, which influence their spectral properties. In contrast to these derivatives, $(MeProp)_2$-Py and $(PhProp)_2$-Py contain alkyl linkers in their ‘arm-like’ substituents, which lead to faster-averaging rotamer landscapes. Therefore, the linker identity has a large impact on the electronic transitions, quantum yields, and lifetimes (vide infra).

Absorption behavior in pyrene derivatives is finely tuned by factors such as molecular symmetry, the nature of substituents, their positions on the pyrene core, and even the type of linkers in the ‘arm-like’ structures. Notably, subtle modifications to the pyrene core can lead to significant differences in photophysical properties. Therefore, precise molecular design is essential to achieve desirable luminescent characteristics.

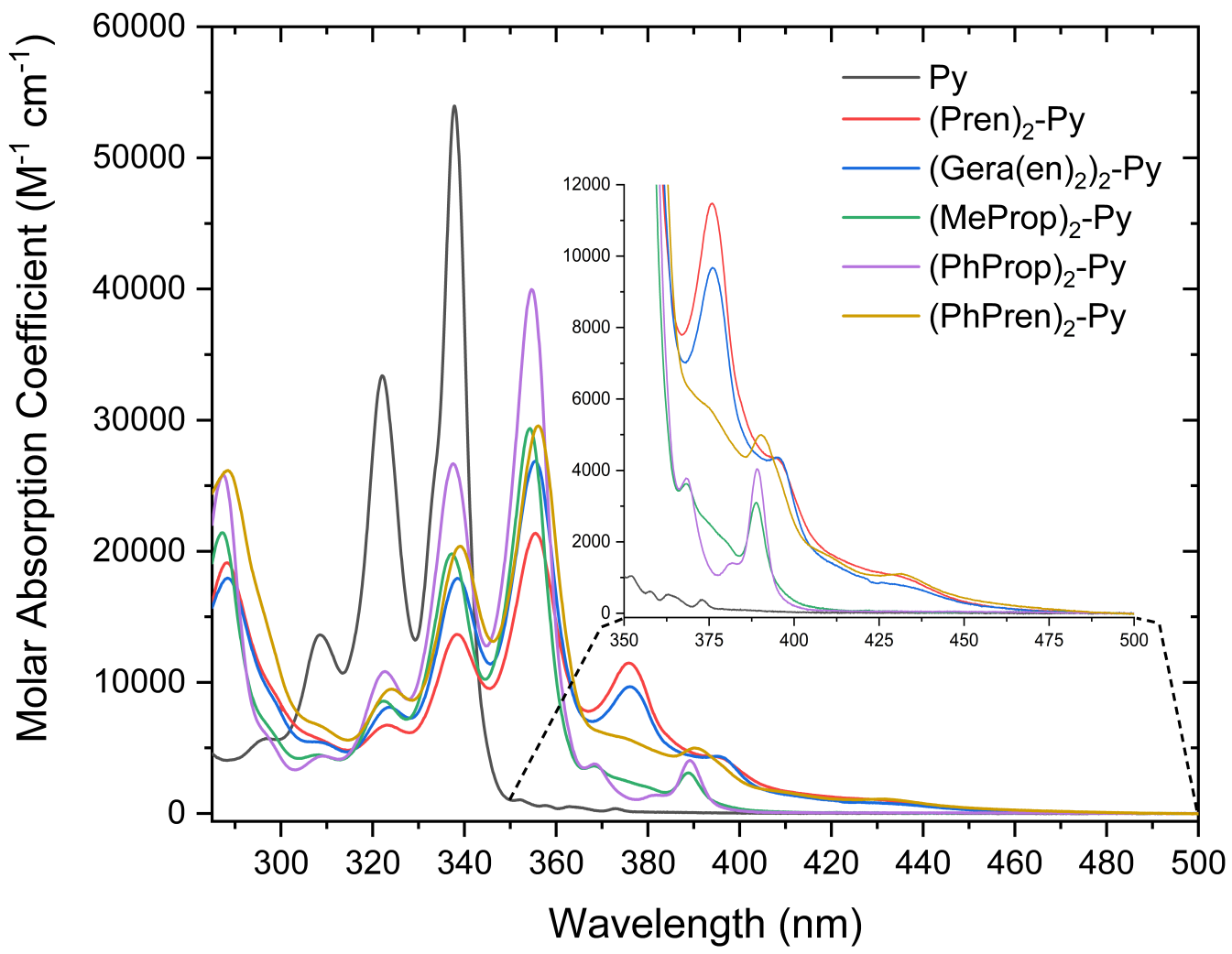

**Figure 2.** UV-visible absorption spectra of dilute, 10 µM, Py (black), (Pren)$_2$-Py (red), (Gera(en)$_2$)$_2$-Py (blue), (MeProp)$_2$-Py (green), (PhProp)$_2$-Py (purple), and (PhPren)$_2$-Py (brown) in toluene in an aerated environment at room temperature. The inset is an expanded view of the transition from the electronic ground state ($S_0$) to the electronic first excited state ($S_1$).

**Table 1. Steady-state Spectroscopic Characteristics of Pyrene and its Derivatives in Toluene**

| Sample | $\lambda_{S_0 \rightarrow S_1}$ [a] (nm) | ε (M$^{-1}$ cm$^{-1}$) | $\lambda_{S_0 \rightarrow S_2}$ [b] (nm) | ε (M$^{-1}$ cm$^{-1}$) | $\lambda_{em}$ [c] (nm) |
|---|---|---|---|---|---|
| Py | 372.9[a] | $3.0 \times 10^2$ | 337.8 | $5.2 \times 10^4$ | 374 |
| (Pren)$_2$-Py | 427.3[a] | $1.4 \times 10^3$ | 355.4 | $2.2 \times 10^4$ | 399 |
| (Gera(en)$_2$)$_2$-Py | 427.4[a] | $1.0 \times 10^3$ | 355.2 | $2.7 \times 10^4$ | 399 |
| (MeProp)$_2$-Py | 427.5[a] | $3.2 \times 10^2$ | 354.0 | $3.0 \times 10^4$ | 391 |
| (PhProp)$_2$-Py | 427.6[a] | $1.4 \times 10^2$ | 354.3 | $4.0 \times 10^4$ | 391 |
| (PhPren)$_2$-Py | 429.7[a] | $1.2 \times 10^3$ | 355.9 | $3.1 \times 10^4$ | 395 |

[a]Wavelength of the origin (0-0) transition in the absorption band, determined to the best of our ability. [b]Wavelength corresponding to the maximum absorbance intensity in the absorption band. [c]Wavelength of the emission maximum.

**Fluorescence Spectroscopy.** To examine the effect of solvent on the fluorescence emission, the derivatives were measured in a series of solvents with varying polarities. This emission behaviour reinforces the conclusion that no significant charge-transfer or π–π stacking interactions occur between the 'arm-like' pyrene derivatives and toluene, consistent with the trends observed in the absorption spectra; see Figure S1B in the SI for the representative case of $(Gera(en)_2)_2$-Py. Figure S4 in the SI demonstrates that the fluorescence excitation spectra of dilute samples reproduce all features of the absorption spectra. This overlap confirms that the fluorescence originates from the same molecular species and excited state as the absorption. It further supports that pyrene and its derivatives are pure and photo-physically stable compounds that obey Kasha's rule, with no evidence of complex photophysical processes.

Figure 3A and 3B emphasize the effect of the freeze–pump–thaw degassing procedure. Dissolved molecular oxygen induces substantial fluorescence quenching in unsubstituted pyrene, whereas the 'arm-like' pyrene derivatives are significantly less affected. In all samples there are minimal spectral shifts. To quantify the fluorescence intensity differences, their quenching effects are evaluated by comparing their fluorescence quantum yields under aerated and degassed conditions. The fluorescence quantum yield under degassed conditions for unsubstituted pyrene is ca. 17 times larger than that under aerated conditions. In comparison, all 'arm-like' derivatives have fluorescence quantum yields of ca. 2 – 3 times larger after deaeration. This is due to two effects. First, is the so-called kinetic-window effect, whereby reduced oxygen encounters are the result of the shorter excited-state lifetimes of the pyrene derivatives (see Table 2). The second effect is the stretched-out 'arm-like' structures at the 1,3-positions acting as protecting groups and encapsulating their pyrene cores physically, ensuring negligible interactions with the dissolved molecular oxygen.

We can determine the most important effect by invoking quenching via the Stern-Volmer relationship.[38] For dilute solutions, collisional quenching is the predominant quenching mechanism.[38,39] Therefore, the dynamic Stern-Volmer equation is appropriate, i.e. $\frac{\Phi_0}{\Phi} = 1 + k_q \tau_0 [O_2]$, where $\Phi_0$ and $\Phi$ represent the quantum yields in the absence and presence of quencher respectively, $k_q$ is the bimolecular quenching rate constant, $\tau_0$ is the lifetime in the absence of quencher (degassed lifetime), and $[O_2]$ is the concentration of the quencher molecules oxygen. If the quenching rate constant and the oxygen concentration are approximately the same for all the samples, the quenching factor (17 for pyrene) scales with the degassed lifetime, i.e. $k_q [O_2] = \frac{\Phi_0/\Phi - 1}{\tau_0} = \frac{16}{332\ \mathrm{ns}} \sim 4.8 \times 10^7\ \mathrm{s}^{-1}$. Assuming that $[O_2]$ is 1.8 – 1.9 mM in toluene at room temperature,[39,40] $k_q$ is approximately $3 \times 10^{10}\ \mathrm{M}^{-1}\mathrm{s}^{-1}$. This value is near the diffusion limit in a low-viscosity solvent, such as toluene, and its magnitude matches classic measurements for oxygen quenching of aromatic singlets in nonpolar solvents.[41] This demonstrates that our fluorescence spectra are accurate, as is the 17 times decrease of pyrene's quantum yield in aerated solution. The quenching factors for the derivatives can be approximated using $1 + k_q \tau_0 [O_2]$ and the degassed lifetimes from Table 2. In the case of a bi-exponential decay, the amplitude-weighted lifetimes are used. This yields a fluorescence quantum yield decrease of 2 – 3 times in aerated solution, compared to degassed solution. This implies that a shorter degassed lifetime leaves less time for dissolved oxygen to collide with the derivatives, which yields a smaller Stern-Volmer response. The quenching rate constant ($k_q$) need not change for this effect to occur. Therefore, the 'arms' contribute a small transient shielding effect, and their main importance is in the lifetime decrease of the derivatives.

Compared to unsubstituted pyrene, the 'arm-like' derivatives have similar emission behaviours with broader and less structured emission $S_1 \rightarrow S_0$ bands. It highlights the influence of substitution on the pyrene core, leading to less vibronic couplings and unique excited-state dynamics. Just as discussed in the Absorption Spectroscopy section, the fluorescence maxima for examined 'arm-like' derivatives in toluene are redshifted by approximately 1160 – 1680 $cm^{-1}$ due to the electron density contribution by 'arm-like' substituents and the electron-donating nature of the hydroxyl group. In agreement with our observations, the presence of alkenyl linkers in the 'arm-like' substituents leads to a more pronounced bathochromic shift. Those derivatives with alkyl linkers in their arms are less redshifted, resulting from reduced π-conjugation, restricting the electronic delocalization from the 'arm-like' substituents to the pyrene core.

The redshifts in the absorption and fluorescence spectra were further quantified to obtain the ground and excited state energy gaps. The energy gap between the $S_0$ and $S_1$ states is ca. 23000 $cm^{-1}$ for all the derivatives, whereas unsubstituted pyrene exhibits a larger gap of ca. 27000 $cm^{-1}$. The energy gap between the $S_0$ and $S_2$ states for all 'arm-like' derivatives is similar, i.e. ca. 28000 $cm^{-1}$. The unsubstituted pyrene has a $S_0 \rightarrow S_2$ band gap of ca. 30000 $cm^{-1}$. Detailed energy levels are recapped in Table S1 in the SI. As a consequence of the electron density donation from the 1,2,3,7-position substituents, compressed energy spacing in the $S_0 \rightarrow S_1$ and $S_0 \rightarrow S_2$ transitions are commonly observed. Of particular importance, the types of linkers in the 'arm-like' substituents have nearly no impact on the $S_0$ and $S_2$, and the $S_0$ and $S_1$ energy gaps. The different linkers are far removed from the pyrene core, so their chemical differences only have a small effect on the electronic coupling and the concomitant transition mixing.

As demonstrated in Figure 4 and Figure S5 in the SI, no redshifted broad structureless band was identified in the fluorescence emission of the derivatives at high concentrations. Therefore, no excimers form in these samples. Numerous previous studies have demonstrated that bulky or sterically hindering substituents on pyrene can suppress excimer formation. This suppression is attributed to impeded and incomplete overlapping π-planes.[2,3,21,42–46] The 2-position hydroxyl group and the 7-position *tert*-butyl group are bulky, sterically suppressing the excimer formation by increasing the closest-approach distance, which is enhanced by the addition of 'arms'. Precisely for this reason, excimer formation is severely hindered for these 'arm-like' pyrene derivatives in solution. Shorter intrinsic lifetimes lead to a reduced formation of excimers, as described in the oxygen quenching argument.

**A**

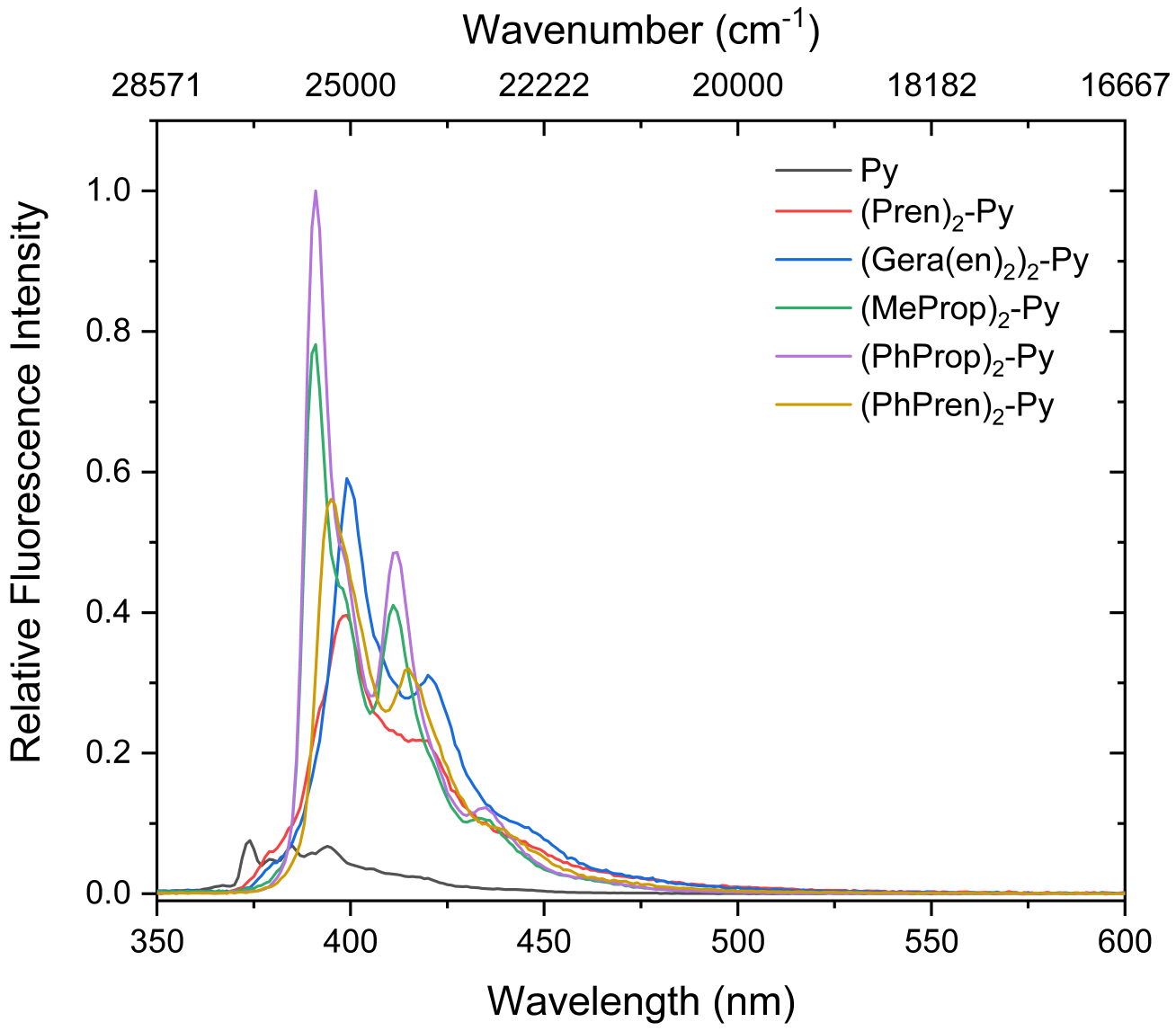

**B**

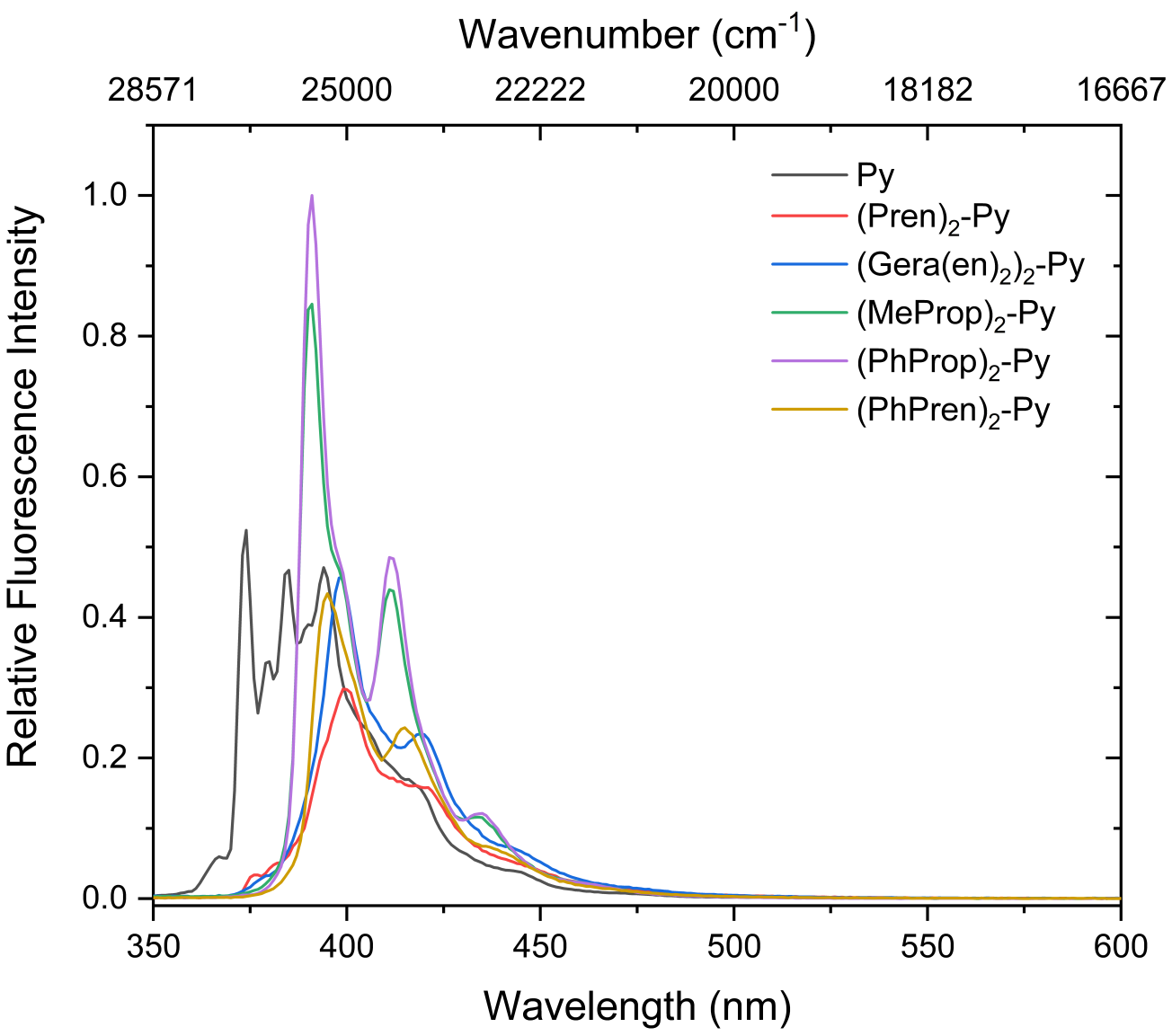

**Figure 3.** Comparison of the relative emission intensities of dilute pyrene and synthesized pyrene derivatives in aerated and degassed toluene at room temperature. Excitation is at 330 nm. (A) Fluorescence emission spectra of Py (black), (Pren)$_2$-Py (red), (Gera(en)$_2$)$_2$-Py (blue), (MeProp)$_2$-Py (green), (PhProp)$_2$-Py (purple), and (PhPren)$_2$-Py (brown) in aerated toluene. (B) Fluorescence emission spectra of Py (black), (Pren)$_2$-Py (red), (Gera(en)$_2$)$_2$-Py (blue), (MeProp)$_2$-Py (green), (PhProp)$_2$-Py (purple), and (PhPren)$_2$-Py (brown) in degassed toluene.

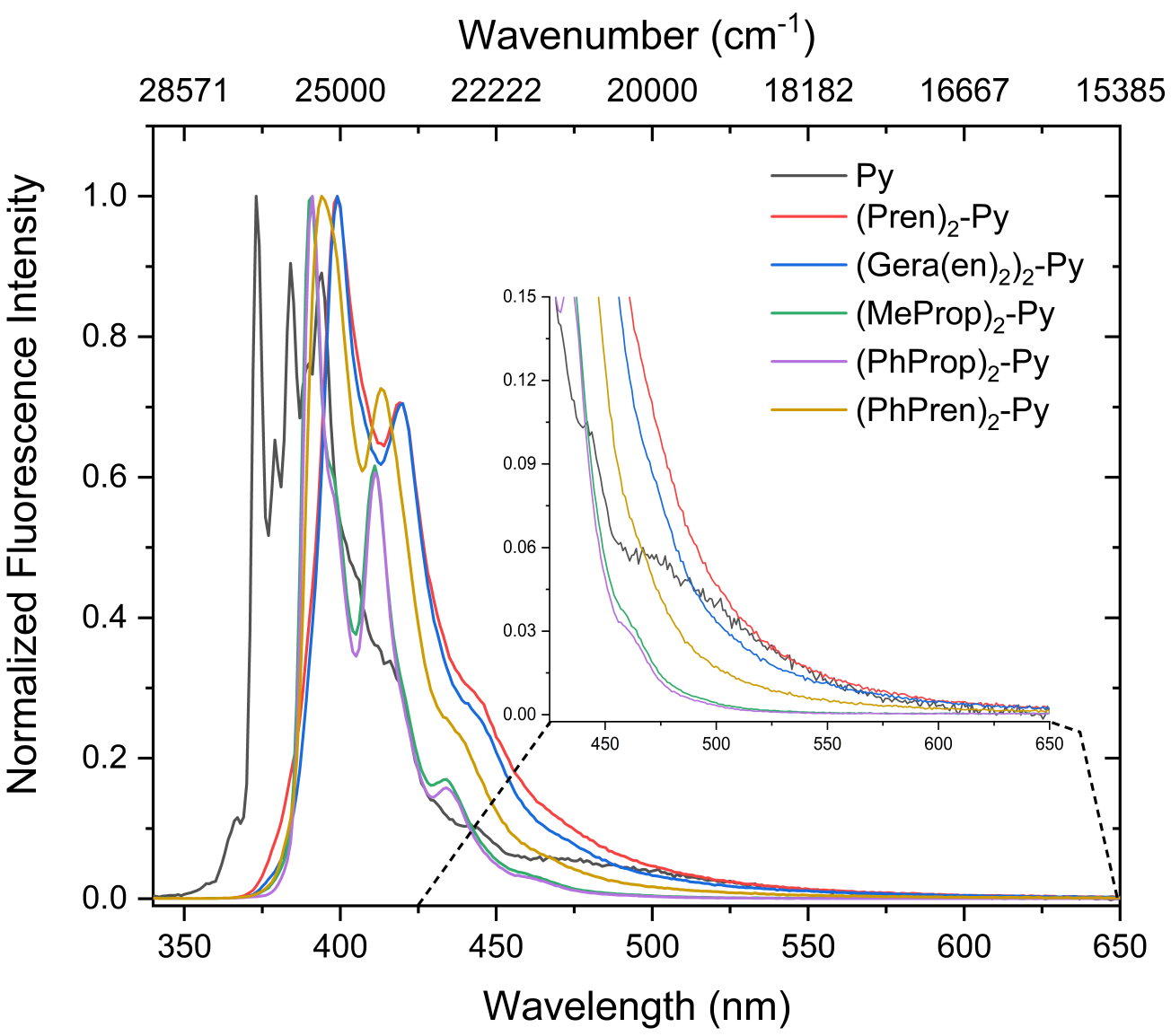

**Figure 4.** Normalized fluorescence emission spectra of Py (black), $(Pren)_2$-Py (red), $(Gera(en)_2)_2$-Py (blue), $(MeProp)_2$-Py (green), $(PhProp)_2$-Py (purple), and $(PhPren)_2$-Py (brown) in toluene at concentrations of 250 μM in an aerated environment at room temperature. Excitation is at 330 nm. Each spectrum was normalized at its maximum intensity. The inset shows the relative intensities of excimer formation in pyrene and its derivatives.

**Fluorescence Quantum Yields and Time-resolved Fluorescence Spectroscopy.** The fluorescence quantum yield of pyrene in deaerated cyclohexane is well-reported, and it ranges from 0.57 to 0.60.[47,48] The fluorescence emission spectra of pyrene in degassed toluene and cyclohexane are shown in Figure S6 in the SI. Taking into account the differences of the solvent refractive indices at the excitation wavelength and the integrated fluorescence emission areas, the fluorescence quantum yield of pyrene in degassed toluene is determined to be 0.65 ± 0.01, using a quantum yield of 0.58 for pyrene in degassed cyclohexane. An earlier study reported the fluorescence quantum yield of pyrene in degassed toluene as 0.64 using the integrating sphere method, which is consistent with the calculated fluorescence quantum yield in this work.[21] Fluorescence quantum yields of 'arm-like' derivatives are determined using pyrene in degassed toluene as a quantum yield standard. The quantum yields and kinetic parameters are summarized in Table 2.

The fluorescence decay profiles are presented in Figure 5 and Figure S7 in the SI. Pyrene has a long-lived fluorescence in degassed toluene of 332 ± 12 ns with a radiative decay constant of ca. $10^6$ $s^{-1}$, which is on the same magnitude as that documented in previous studies.[21] The larger radiative rate constant for pyrene indicates fluorescence emission is the dominant energy decay pathway. The lower nonradiative decay rate constant for pyrene shows that other decay pathways, including internal conversion, vibrational relaxation, as well as intersystem crossing ($10^5$ $s^{-1}$)[21] are uncompetitive. Pyrene derivatives have shorter fluorescence lifetimes and mostly lower quantum yields of 0.4 – 0.7 compared to pyrene. This is because of their less forbidden $S_0$ $\rightarrow S_1$ transitions, which leads to more nonradiative decay pathways competing efficiently with radiative decay. Predictably, a reduction in fluorescence lifetimes is observed for pyrene at higher concentrations, where excimer formation is aided by strong π–π stacking interactions. In contrast, the 'arm-like' pyrene derivatives exhibit no significant variation in fluorescence lifetimes across the studied concentrations. Instead, the unaltered lifetimes of 'arm-like' pyrene derivatives indicate the absence of excimer formation in these systems.

The shorter derivative lifetimes, with a concomitant larger radiative decay rate ($k_r$), compared to pyrene, are clearly shown in Table 2. To confirm that these changes are due to symmetry-breaking, we estimated radiative lifetimes ($\tau_{rad}$) from absorption spectra integrals using the Strickler-Berg (SB) equation.[49] Using $\tau_{rad}$ of pyrene of 500 ns, the $\tau_{rad}$ for the alkenyl-chain-containing derivatives, such as $(Pren)_2$-Py, $(Gera(en)_2)_2$-Py, were found to be 65-70 ns; for the alkyl-chain-containing derivatives $(MeProp)_2$-Py and $(PhProp)_2$-Py, 95-105 ns. The electric-dipole allowed assumption implicit in the SB equation leads to differences between the radiative rate ($k_r$) calculated with the SB equation and $k_r$ calculated via $\Phi/\tau$ given in Table 2. However, the correct order of magnitude of $k_r$ for the derivatives is reproduced. This demonstrates that the higher $k_r$ for the derivatives, compared to pyrene, is based on the symmetry breaking effect due to the sterically bulky OH and *tert*-butyl groups and the 'arms' that promote coupling of the transitions. The differences in the $\tau_{rad}$ magnitudes calculated for the different chains – ca. 67 ns for the alkenyl linker derivatives and ca. 100 ns for the alkyl linker derivatives – reveal that the $S_0$-$S_1$ transition is approximately 1.5 times brighter. This larger oscillator strength for the alkenyls can be clearly seen in Figure 2. The inference is that some extra intensity borrowing into $S_1$ occurs for the alkenyl derivatives.

In Table 2, the fluorescence lifetimes of $(MeProp)_2$-Py and $(PhProp)_2$-Py, containing alkyl linkers, are shown to have a relatively longer-lived fluorescence, ca. 40 ns. On the other hand, $(Pren)_2$-Py, $(Gera(en)_2)_2$-Py, and $(PhPren)_2$-Py, containing alkenyl linkers, have a relatively shorter-lived fluorescence ranging between 20 – 30 ns. $(Pren)_2$-Py, $(Gera(en)_2)_2$-Py also display bi-exponential excited-state fluorescence decays at low (Table 2) and high (Table S2 in the SI) concentrations. These time-resolved measurements are in excellent agreement with their observed fluorescence quantum yields. Since $\tau_{rad}$ values calculated from the SB equation are similar for the derivatives, but the lifetimes and quantum yields vary, this effect must come from the nonradiative rate ($k_{nr}$). It was already mentioned above that $k_{nr}$ for pyrene is an order of magnitude smaller than for the derivatives due to the derivatives having less forbidden $S_0 \rightarrow S_1$ transitions with extra nonradiative decay pathways. When $k_{nr}$ is compared for the derivatives, derivatives with the unsaturated alkenyl linkers almost double $k_{nr}$ compared to those with the saturated alkyl linkers. The sole structural variation among these organic frameworks lies in the nature of the linkers, unsaturated versus saturated. Although the alkenyl linkers are more rigid, they introduce soft, π-coupled motions, such as allylic torsions and C=C out-of-plane wags on the order of 400 $cm^{-1}$. These torsional motions lead to a large vibrational density that couples the $S_1$ and $S_0$ states more efficiently, which increases internal conversion and results in larger $k_{nr}$ constants.

The origin of the bi-exponential lifetimes also arises due to the structure of the alkenyl-arms. Two kinetic subpopulations exist for the alkenyl-linker derivatives. The major population is an open conformer with a slow decay of ca. 30 ns, with a minor contribution (10 – 20%) originating from a back-folded or twisted conformer with a faster decay of ca. 9 ns. The twisted conformer might back-fold in such a way that the alkenyl segment is closer to the pyrene core, which would enhance vibronic coupling and lead to a larger $k_{nr}$ constant. The bi-exponential decay implies that the interconversion between these different conformers is slower than emission so that both decays are resolved. In contrast, the alkyl 'arms' have simpler, faster-averaging rotamer landscapes. Conformers interconvert faster than emission, so a mono-exponential excited-state decay is measured. In short, the type of linkers within the 'arm-like' substituents significantly alter pyrene derivatives' luminescent properties, which enables targeted design and tuning of these organic conjugated molecules.

**Table 2. Fluorescence Quantum Yields, Lifetimes, and Kinetic Parameters for Dilute Pyrene and its Derivatives in Degassed Toluene**

| Sample | $\Phi^{a}_{f}$ | $\tau^{b}_{f}$ (ns) | $k^{c}_{r}$ ($s^{-1}$) | $k^{d}_{nr}$ ($s^{-1}$) |
|---|---|---|---|---|
| Py | $0.65 \pm 0.01$ | $332 \pm 12$ | $2.0 \times 10^{6}$ | $1.1 \times 10^{6}$ |
| $(Pren)_2$-Py | $0.40 \pm 0.06$ | $29.8 \pm 0.3$, $8.1 \pm 0.3$[e] | $1.5 \times 10^{7}$ | $2.3 \times 10^{7}$ |
| $(Gera(en)_2)_2$-Py | $0.48 \pm 0.02$ | $28.3 \pm 0.6$, $8.5 \pm 1.5$[e] | $1.9 \times 10^{7}$ | $2.0 \times 10^{7}$ |
| $(MeProp)_2$-Py | $0.63 \pm 0.03$ | $37.1 \pm 0.4$ | $1.7 \times 10^{7}$ | $1.0 \times 10^{7}$ |
| $(PhProp)_2$-Py | $0.68 \pm 0.03$ | $36.5 \pm 0.2$ | $1.9 \times 10^{7}$ | $8.8 \times 10^{6}$ |
| $(PhPren)_2$-Py | $0.41 \pm 0.01$ | $24.4 \pm 0.3$ | $1.7 \times 10^{7}$ | $2.4 \times 10^{7}$ |

[a]Fluorescence quantum yield. [b]Fluorescence lifetime extracted from fits to the decay profiles in Figure 5. [c]Radiative decay rate constant, calculated as $\Phi_f/\tau_f$. [d]Nonradiative rate constant, calculated as $(1/\tau_f) - k_r$, where $\tau_f = (\tau_1 \times n_1) + (\tau_2 \times n_2)$. n represents the relative % amplitude, and $n_1 + n_2 = 1$. [e]Decay profiles fitted using a two-component exponential model, reported as $\tau_1$ and $\tau_2$.

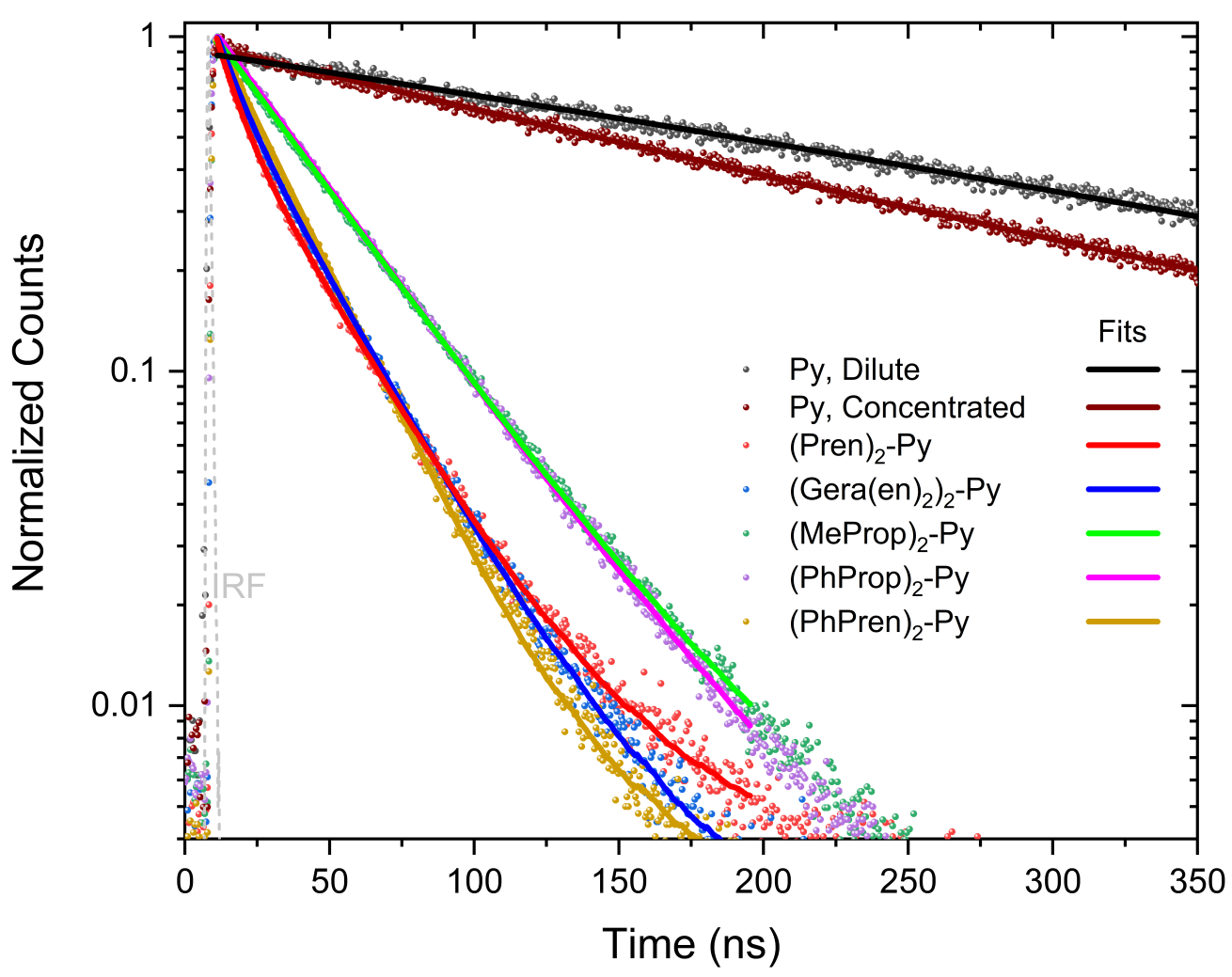

**Figure 5.** Nanosecond time-resolved fluorescence decay profiles of pyrene and its derivatives in degassed toluene at room temperature. Excitation at 337 nm; detection at Py, Dilute (black, 370 nm), Py, Concentrated (wine, 370 nm), $(Pren)_2$-Py (red, 395 nm), $(Gera(en)_2)_2$-Py (blue, 395 nm), $(MeProp)_2$-Py (green, 387 nm), $(PhProp)_2$-Py (purple, 387 nm), and $(PhPren)_2$-Py (brown, 390 nm).

## CONCLUSIONS

The photophysical properties of pyrene and novel 'arm-like' derivatives were thoroughly investigated using optical and time-resolved laser spectroscopy techniques. The investigations first focused on the photophysical characterization of pyrene in toluene, which addresses a current gap in the literature while also establishing reliable benchmarks for our further studies. The $S_0 \rightarrow S_1$ transition in pyrene is symmetry-forbidden by the Laporte rule, where the reported pyrene derivative with substitutions at the 1,2,3,7-positions break the symmetry, which increases the allowedness of the $S_0 \rightarrow S_1$ transitions. In toluene, it was demonstrated that pyrene maintains its propensity for excimer formation at higher concentrations (ca. 30 μM), whereas the 'arm-like' pyrene derivative exhibit no excimer formation. This is attributed to the presence of both the hydroxyl and *tert*-butyl, at the 2- and 7-positions respectively, and the additional bulky 'arm-like' substituents at the 1- and 3-positions. This is confirmed by steady-state fluorescence emission and time-resolved decay profiles. The shorter lifetimes of the pyrene derivatives lead to a reduction of the susceptibility to energy scavenging by molecular oxygen. Also, alkyl-containing and alkenyl-containing arms have different decay pathways. This finding has the potential to influence the design of pyrene-based materials to increase their stability and performance. Pyrene has a 332 ± 12 ns fluorescence lifetime in deaerated toluene, with a reported fluorescence quantum yield of 0.65 ± 0.01. Comparatively, the 'arm-like' pyrene derivatives have moderate (0.4) to high (0.7) fluorescence quantum yields in deoxygenated toluene. These pyrene derivatives display shorter fluorescence lifetimes from ca. 20 to 40 ns. In general, this study reported the photophysical properties of novel 1,3-substituted 7-*tert*-butylpyren-2-ol derivatives and suggested new synthetic strategies for pyrene derivatives to prevent excimer formation and mitigate the issue of oxygen-induced fluorescence quenching. Due to their exceptional photophysical properties, these novel compounds are of significant interest and are expected to find widespread applications in electronic and optoelectronic fields.

## ASSOCIATED CONTENT

**Supporting Information**

The Supporting Information is available free of charge on the ACS Publications website at DOI: 10.1021/acs.jpca.XXXXXXX.

Synthetic procedures for all pyrene derivatives; $^1$H/$^{13}$C NMR spectra; UV–vis absorption and fluorescence excitation and emission spectra; solvent polarity effects; photodegradation studies; time-resolved decay profiles; tables of energy gaps and fluorescence lifetimes (PDF)

## AUTHOR INFORMATION

**Corresponding Author**

**Amy L. Stevens** – *Department of Chemistry, University of Saskatchewan, Saskatoon, Saskatchewan S7N 5C9, Canada*; orcid.org/0000-0002-0998-1795; Email: amy.stevens@usask.ca

**Authors**

**Wenlong Li** – *Department of Chemistry, University of Saskatchewan, Saskatoon, Saskatchewan S7N 5C9, Canada*

**Stephen Awuku** – *Department of Chemistry, University of Saskatchewan, Saskatoon, Saskatchewan S7N 5C9, Canada*

**Jenna N. Merk** – *Department of Chemistry and Biochemistry, University of Regina, Regina, Saskatchewan S4S 0A2, Canada*

**Marc R. MacKinnon** – *Department of Chemistry and Biochemistry, University of Regina, Regina, Saskatchewan S4S 0A2, Canada*

Completed contact information is available at:
https://pubs.acs.org/doi/10.1021/acs.jpca.XXXXXXX

**Notes**

The authors declare no competing financial interest.

## ACKNOWLEDGEMENTS

The authors are pleased to acknowledge the Natural Sciences and Engineering Research Council of Canada for its continuing financial support (RGPIN-2021-03865). The Saskatchewan Structural Science Centre (SSSC) provided equipment and technical assistance. W.L., S.A., and A.L.S. thank the University of Saskatchewan, the College of Arts and Science, and the Centre for Quantum Topology and Its Applications (quanTA) for partial financial support. W.L. and S.A. acknowledge the College of Graduate and Postdoctoral Studies for scholarship funding. W.L. acknowledges Dr. Marcelo Sales for the use of the melting point apparatus. W.L. and S.A. acknowledge Dr. Valerie J. MacKenzie for the use of the Cary-3500 multicell UV-vis spectrophotometer. W.L., S.A., and A.L.S. acknowledge Chuyao Han's contribution to the design of the graphical abstract. J.N.M. and M.R.M. thank the University of Regina, the Faculty of Science and Department of Chemistry and Biochemistry for partial financial support. J.N.M. acknowledges the Faculty of Graduate Studies (Regina) for scholarship funding.